\documentclass[amsfonts,amsmath]{PoS}
\usepackage{amsfonts,amsmath}

\graphicspath{{./Figs/}}

\newcommand{\beq}{\begin{equation}}
\newcommand{\eeq}{\end{equation}}
\newcommand{\bea}{\begin{eqnarray}}
\newcommand{\eea}{\end{eqnarray}}

\newcommand{\Fig}[1]{Fig.~\ref{#1}}

\title{\begin{center} Landau gauge gluon and ghost propagators from 
two-flavor lattice QCD at $T>0$ \end{center}}

\ShortTitle{}

\author{R. Aouane, F. Burger, \speaker{M. M\"uller-Preussker}, \\
        Humboldt-Universit\"at zu Berlin, Institut f\"ur Physik, 
        12489 Berlin, Germany
}
\author{E.-M. Ilgenfritz\\
       Joint Institute for Nuclear Research, VBLHEP, 141980 Dubna, Russia}
\author{A. Sternbeck \\
       Universit\"at Regensburg, Institut f\"ur Theoretische Physik, 
       93040 Regensburg, Germany}

\abstract{
In this contribution we extend our unquenched computation of the 
Landau gauge gluon and ghost propagators in lattice QCD at non-zero 
temperature. The study was aimed at providing input for investigations 
employing continuum functional methods. We show data which correspond 
to pion mass values between $300$ and $500~\mathrm{MeV}$ and are obtained 
for a lattice size $32^3 \times 12$. 
The longitudinal and transversal components of the 
gluon propagator turn out to change smoothly through the crossover region, 
while the ghost propagator exhibits only a very weak temperature 
dependence. For a pion mass of around $400~\mathrm{MeV}$ and the intermediate
temperature value $T \simeq 240~\mathrm{MeV}$ we compare our results with
additional data obtained on a lattice with smaller Euclidean time extent 
$N_\tau=8,10$ and find a reasonable scaling behavior.
}

\FullConference{
The 31 International Symposium on Lattice Field Theory - Lattice 2013, \\
July 29 - August 3, 2013 \\
Mainz, Germany}

\begin{document}

\section{Introduction}

Over the last 15 years Dyson-Schwinger (DS) equations
and functional renormalization group (FRG) equations
turned into powerful schemes to address various non-perturbative 
problems in QCD. Within this framework 
(for reviews see e.g. \cite{Fischer:2008uz,Boucaud:2011ug}) 
the Landau gauge gluon and ghost propagators appear -- together with the 
corresponding vertices -- as the main building blocks.
In order to solve the (infinite) tower of corresponding equations an 
appropriate truncation has to be applied. Then independent information,  
preferably from first principles, is welcome to improve the 
resulting approximations or to be used as an input to the system of 
equations. 

Lattice QCD calculations allow to provide e.g. the Landau gauge 
gluon and ghost propagators in an ab-initio way. The available momentum 
range, however, is restricted from above by the lattice spacing and 
from below by the available lattice volume up to a further uncertainty 
related to so-called Gribov copies (see e.g. \cite{Sternbeck:2012mf} 
and references therein).

In recent years the DS / FRG framework has been applied also to
explore the non-zero temperature regime of QCD and even 
the non-zero chemical potential case (see \cite{Fischer:2012vc} 
and references therein).

Correspondingly we have started a lattice computation of gluon
and ghost propagators at non-zero temperature, first in 
$SU(3)$ pure gauge theory \cite{Aouane:2011fv} and more recently 
for the full QCD case with two light flavor degrees of freedom  
\cite{Aouane:2012bk}. Here we report on the latter case. For that we
could rely on investigations of the QCD thermodynamics employing 
the Wilson-twisted mass discretization at maximal twist     
\cite{Burger:2011zc,Burger:2012zz}. For three pion mass values 
smooth crossover regimes were found, where the signals for the
transition taken from the (renormalized) Polyakov loop and 
from the chiral condensate (susceptibility) occured at different 
temperature values $T_\mathrm{deconf}$ and $T_\chi$, respectively.

Throughout those crossover regions we have obtained the momentum 
dependence of the transverse und longitudinal components of the 
Landau gauge gluon propagator as well as of the corresponding 
ghost propagator. Additionally to the results published in 
\cite{Aouane:2012bk} we add a first scaling test in order to 
check, in as far we are close to the continuum limit in the 
momentum range considered.

\section{Observables} 

Our first quantity of interest is the gluon propagator in momentum space
\beq
 D^{ab}_{\mu\nu}(q) 
 =\left\langle \widetilde{A}^a_{\mu}(k)\widetilde{A}^b_{\nu}(-k) \right\rangle\,,
\label{eq:gluonzero}
\eeq
where $\langle\cdots\rangle$ represents the path integral average estimated
by averaging over gauge-fixed (Hybrid) Monte Carlo generated gauge field 
configurations. $\widetilde{A}^a_{\mu}(k)$ denotes the Fourier transform of the 
gauge potential. $k_\mu\in \left(-N_{\mu}/2, N_{\mu}/2\right]$ is the
lattice momentum ($\mu=1,\ldots,4$) related to the physical momentum via
$q_{\mu}(k_{\mu}) = \frac{2}{a} \sin\left(\frac{\pi k_{\mu}}{N_{\mu}}\right)$. 
We use the notation $(N_{\sigma}; N_{\tau}) \equiv(N_i;N_4)$ where $i=1,2,3$.

The gauge is fixed with the help of a {\it simulated annealing} prescription
to extremize the Landau gauge functional always followed by overrelaxation 
steps until a prescribed numerical precision for the local gauge condition
is reached. For a discussion about the efficiency of the simulated annealing
procedure in Landau gauge fixing see \cite{Bogolubsky:2007pq}.
 
For non-zero temperature the Euclidean invariance is broken. Therefore, it is 
useful to split $D^{ab}_{\mu\nu}(q)$ into two components, the transversal $D_T$
(``chromomagnetic'') and the longitudinal $D_L$ (``chromoelectric'') propagator,
respectively. For $D_{T,L}$ [or their respective dimensionless 
dressing functions  $Z_{T,L}(q)=q^2 D_{T,L}(q)$] one finds 
(see also \cite{Aouane:2011fv})
\beq
 D_T(q)=\frac{1}{2 N_g} 
        \left\langle \sum_{i=1}^3  
         \widetilde{A}^a_i(k) \widetilde{A}^a_i(-k)
        -\frac{q_4^2}{\vec{q}^{\;2}} 
        \widetilde{A}^a_4(k)\widetilde{A}^a_4(-k)\right\rangle
\eeq
and
\beq
 D_L(q)= \frac{1}{N_g}\left(1 + \frac{q_4^2}{\vec{q}^{\;2}}\right) 
        \left\langle \widetilde{A}^a_4(k) \widetilde{A}^a_4(-k) \right\rangle
\; ,
\eeq
where $N_g=N_c^2-1$ and $N_c=3$.

The zero-momentum gluon propagator values can be defined as
\bea
\label{eq:zeromomprop}
D_T(0) &=& \frac{1}{3 N_g}
     \sum_{i=1}^3 \left\langle \widetilde{A}^a_i(0) \widetilde{A}^a_i(0)
\right\rangle, 
\\
D_L(0) &=& \frac{1}{N_g}
       \left\langle \widetilde{A}^a_4(0) \widetilde{A}^a_4(0) \right\rangle \;. 
\eea

The Landau gauge ghost propagator is given by
\bea
\label{eq:ghost} \nonumber
G^{ab}(q)&=&a^{2}\sum_{x,y}\langle e^{-2\pi i(k/N)\cdot(x-y)} [M^{-1}]^{ab}_{xy}\rangle \\
         &=&\delta^{ab}~G(q)= \delta^{ab}~J(q)/q^2\;,
\eea
where $(k/N) \equiv (k_{\mu}/N_{\mu})$. $J(q)$ denotes the ghost
dressing function. The matrix $M$ is the lattice Faddeev-Popov operator, 
for more details see~\cite{Aouane:2011fv}. For the inversion of $M$ we use the
pre-conditioned conjugate gradient algorithm of \cite{Sternbeck:2005tk} with
plane-wave sources $\vec{\psi}_{c}$ with color and position components 
$\psi^{a}_{c}(x)=\delta^{a}_{c}\exp(2\pi\,ik\cdot(x/N))$.

In order to reduce lattice artifacts we restrict ourselves to diagonal 
and slightly off-diagonal momenta for the gluon propagator and diagonal
momenta for the ghost propagator. Moreover, only modes with 
zero Matsubara frequency ($k_4=0$) are considered.

Assuming that the influence of lattice artifacts can be neglected the
renormalized dressing functions, defined in momentum subtraction (MOM) 
schemes, can be obtained via multiplicative renormalization 
\bea
Z_{T,L}^{ren}(q,\mu) &\equiv& \tilde{Z}_{T,L}(\mu) Z_{T,L}(q),  \nonumber \\
J^{ren}(q,\mu) &\equiv& \tilde{Z}_{J}(\mu) J(q) \label{eq:renormZ}
\eea
with the $\tilde{Z}$-factors being defined such that
$Z_{T,L}^{ren}(\mu,\mu) =  J^{ren}(\mu,\mu) = 1$. Our
renormalization scale throughout this paper is $\mu=2.5~\mathrm{GeV}$
(see also Ref. \cite{Aouane:2012bk}).

\section{Results}

Our estimates for $D_{T,L}$ and $G(q,T)$ rely on gauge field ensembles 
generated on a lattice of size $N_\sigma=32$ and $N_\tau=12$ and  
for parameters which are collected in detail in \cite{Aouane:2012bk}.

\begin{figure*}[tb]
 \centering
 \mbox{
 \includegraphics[height=4.0cm,width=12cm]{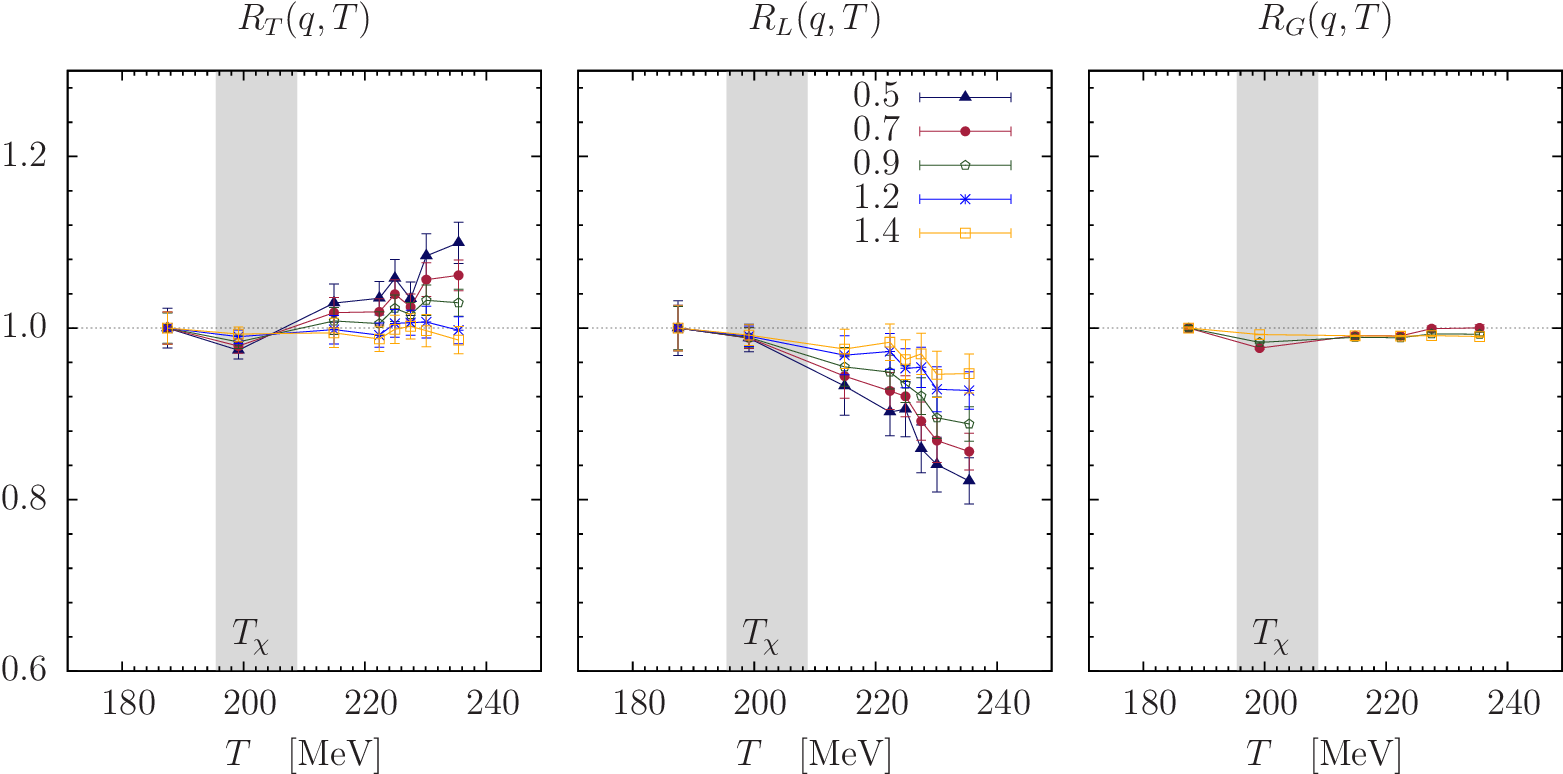} 
 }
 \par\vspace*{0.7cm}
 \centering
 \mbox{
 \includegraphics[height=4.0cm,width=12cm]{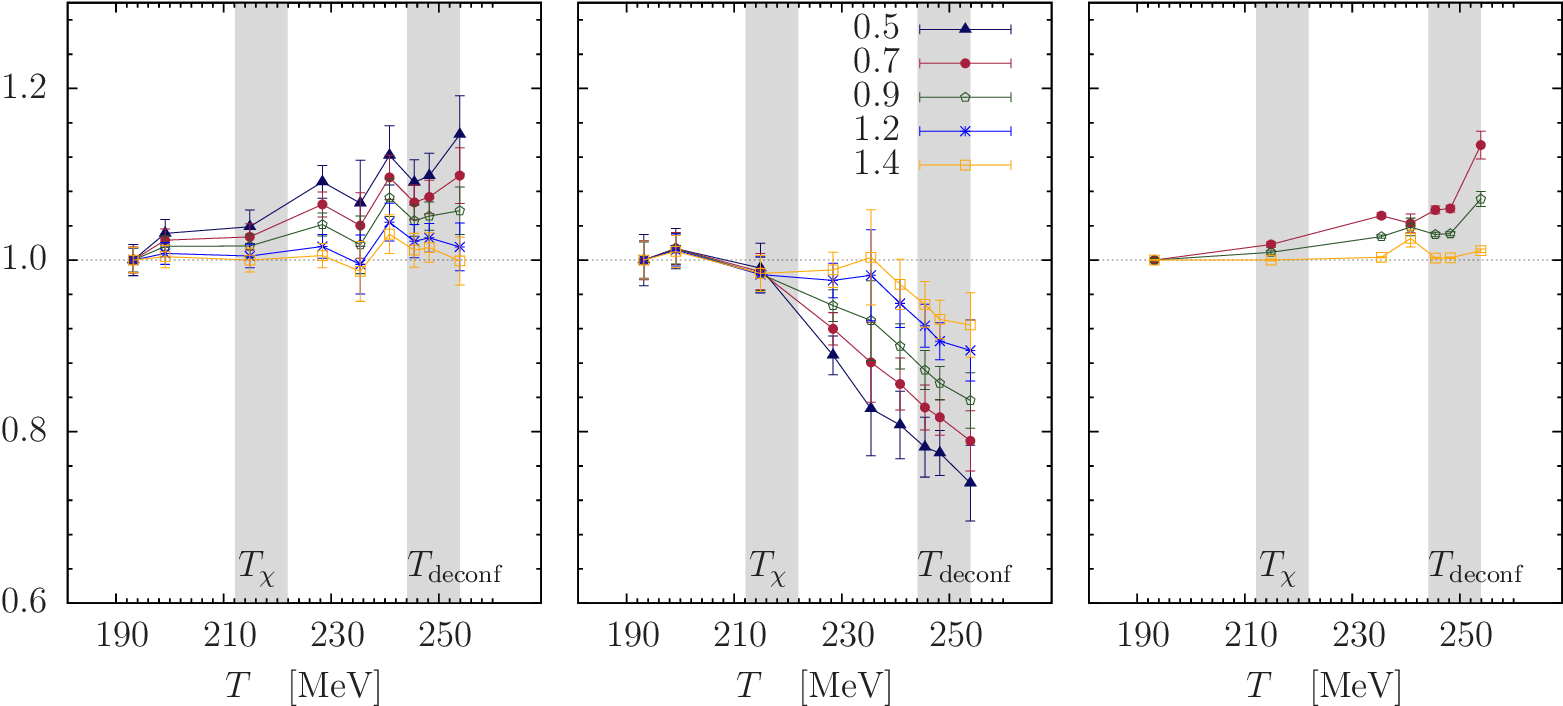} 
 }
 \par\vspace*{0.7cm}
\centering
 \mbox{
 \includegraphics[height=4.0cm,width=12cm]{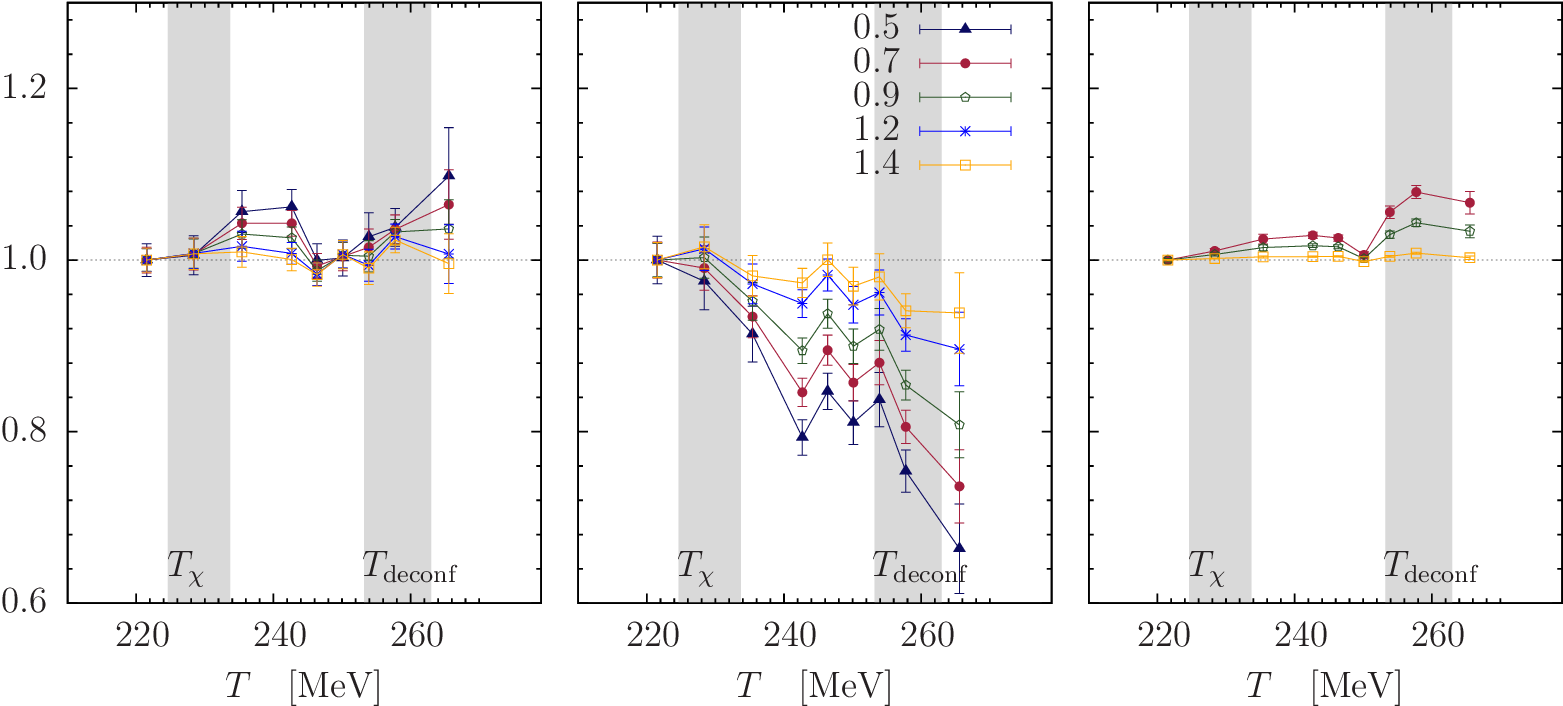} 
 }
\caption{Ratios $R_T, R_L$ and $R_G$ for the renormalized 
transverse $D_{T}^{ren}$ (left panel), longitudinal 
$D_{L}^{ren}$ (middle panel) and ghost $G^{ren}$ (right panel) propagators, 
respectively, as  functions of the temperature $T$ at a few non-zero
momentum values $q$ (indicated in units of $\mathrm{[GeV]}$. The corresponding 
pion masses (from top to bottom) are $m_{\pi} \simeq 316,~398$ and 
$469~\mathrm{MeV}$. The vertical bands indicate the chiral ($T_\chi$)
and `deconfinement' ($T_\mathrm{deconf}$) pseudo-critical temperatures 
with their uncertainties.}  
\label{fig:props_vs_T}
\end{figure*}
In \Fig{fig:props_vs_T} we show ratios of the renormalized dressing 
functions or propagators
\bea
R_{T,L}(q,T) &=& D_{T,L}^{ren}(q,T) / D_{T,L}^{ren}(q,T_\mathrm{min}), \\
R_G(q,T) &=& G^{ren}(q,T)/G^{ren}(q,T_\mathrm{min}) 
\eea
as functions of the temperature $T$ for 6 fixed (interpolated) momentum values 
$q \ne 0$, and for the three different pion masses (panels from top to bottom). 
For better visibility, ratios are normalized with respect to the
respective left-most shown temperature $T_\mathrm{min}$ 
in \Fig{fig:props_vs_T}. 

We see $R_L(q,T)$ to decrease more or less monotonously with the
temperature through the crossover region, and this
decrease is stronger the smaller the momentum is. 
This behavior is similar but much less pronounced than what 
we saw in the close neighborhood of the phase transition for the quenched 
case \cite{Aouane:2011fv}, where a first-order phase transition takes 
place.

$R_T(q,T)$ instead signals a slight increase within the same range, and 
the ghost propagator (at fixed low momenta) seems to rise near
$~T \simeq T_\mathrm{deconf}$.

\begin{figure*}[tb]
 \centering
  \mbox{
 \includegraphics[height=4.5cm,width=14cm]{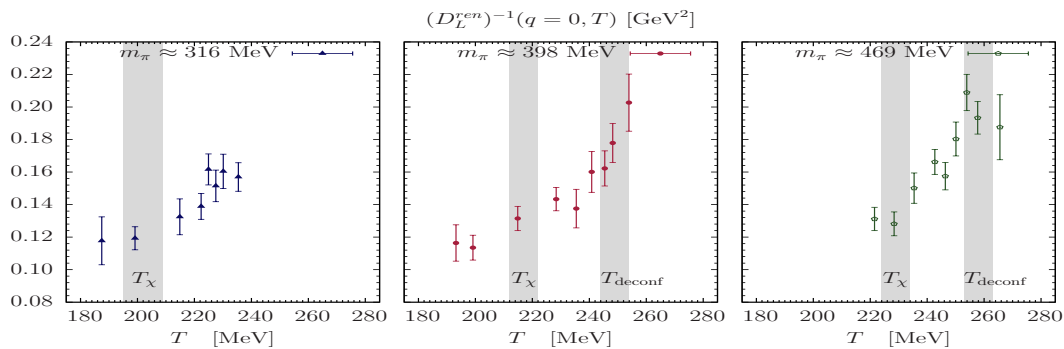} 
 }
\caption{The panels show the inverse renormalized 
longitudinal gluon propagator $(D_L^{ren})^{-1}$ at zero momentum
for the three pion mass values indicated.}
\label{fig:gluon_zero_momentum}
\end{figure*}
The panels of \Fig{fig:gluon_zero_momentum}, show data for the inverse
renormalized longitudinal propagator $1/D^{ren}_L$ at zero momentum, 
again versus temperature and separately for different pion masses.
Assuming a linear $q^2$-behavior as $D^{ren}_L(q,T)^{-1} \sim q^2 + m_{sc}(T)^2$ 
at low $q^2$, this quantity can be related to a {\it gluon screening mass} 
$m_{sc}(T)$. Since $D^{ren}_L(0)^{-1}$ rises with temperature in the 
crossover region it may also serve as a useful indicator for the 
finite-temperature crossover of the quark-gluon system. However, 
such zero-momentum results are always influenced by strong finite-size 
and Gribov copy effects, which we have not analyzed here. 

\begin{figure*}[tb]
 \centering
  \mbox{
 \includegraphics[height=4.5cm,width=6.0cm]{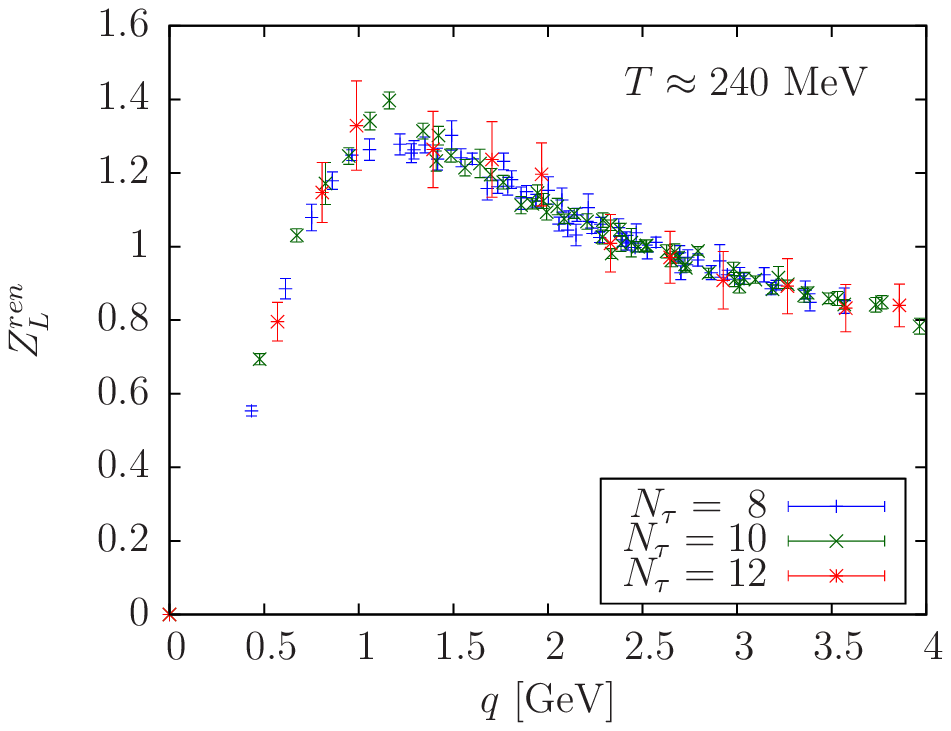} 
 \includegraphics[height=4.5cm,width=6.0cm]{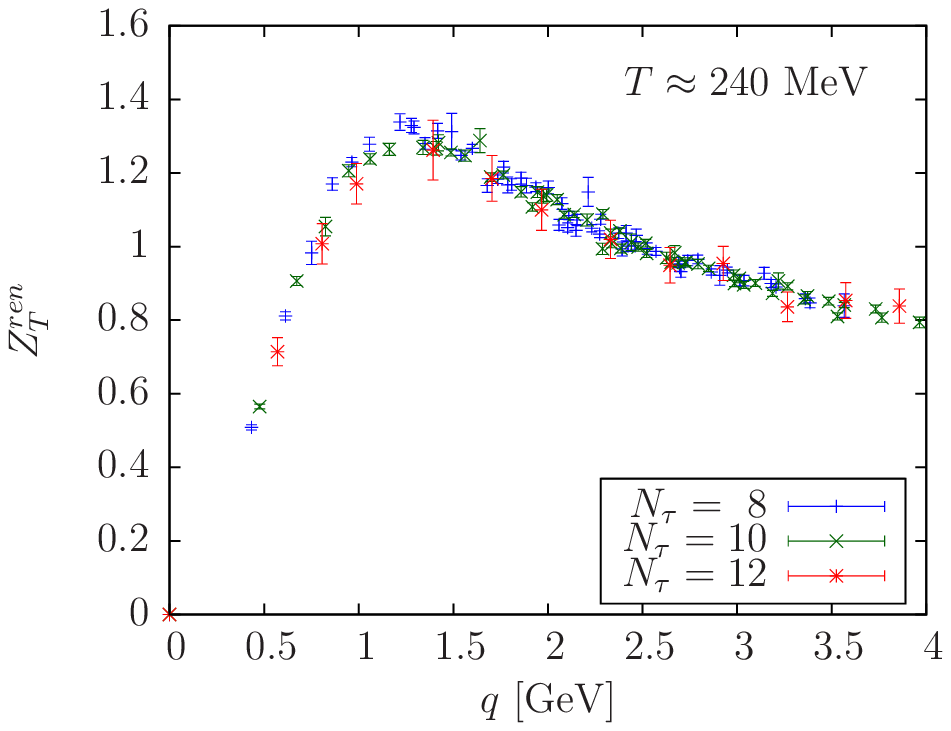} 
 }
\caption{The left (right) panel shows the renormalized 
longitudinal (transversal) gluon dressing function for a
scaling check by varying $N_{\tau}=8, 10, 12$ at 
$m_\pi\simeq 398$ MeV  and $T \simeq 240$ MeV.}
\label{fig:scaling_check}
\end{figure*}
Finally in \Fig{fig:scaling_check} we show a scaling test 
achieved with gauge field ensembles produced additionally
at smaller time-like lattice extent $N_{\tau}$ but at 
approximately the same values of the charged pion mass 
($m_\pi\simeq 398$ MeV) and temperature $T \simeq 240$ MeV 
inside the crossover region.  
We see that the results for the two largest values $N_{\tau}$
nicely agree within the achieved statistical errors telling us
that the $N_{\tau}=12$ results seem already to be close to the
continuum limit. For the latter results, in Ref. \cite{Aouane:2012bk}   
we have provided fit formulae working reasonably well 
in the momentum range $~0.4~\mathrm{GeV} \le q \le 3.0~\mathrm{GeV}$
and hopefully being useful as input for the continuum
DS or FRG framework. 

\vspace*{0.2cm}
We thank the HLRN supercomputing centers Berlin/Hannover for 
providing us with the necessary computing resources. F.B.~and 
M.M.P.~acknowledge support from DFG with an SFB/TR9 grant and 
R.A. from the Yousef Jameel Foundation at Humboldt-University Berlin.
A.S. acknowledges support by the European Reintegration Grant
(FP7-PEOPLE-2009-RG, No.256594).

\bibliographystyle{utphys}
\providecommand{\href}[2]{#2}\begingroup\raggedright\endgroup

\end{document}